\documentstyle[eps,12pt]{article}
\def\reference{\parskip 0pt\par\noindent\hangindent 0.5 truecm}

\newcommand{\simgt}{\raisebox{-0.7ex}{\mbox{$\stackrel{\textstyle >}{\sim}$}}}
\newcommand{\simlt}{\raisebox{-0.7ex}{\mbox{$\stackrel{\textstyle <}{\sim}$}}}

\begin{document}

\title{Obscuration by Diffuse Cosmic Dust}

\author{Frank J. Masci
} 

\date{}
\maketitle

{\center
School of Physics, University of Melbourne, Parkville, Victoria, Australia, 3052\\fmasci@physics.unimelb.edu.au\\[3mm]
}

\begin{abstract}
If the background universe is observed
through a significant amount of diffusely distributed foreground dust, then
studies
at optical wavelengths may be severely biased.
Previous studies investigating the effects of foreground dust
on background sources assumed dust to be ``compactly'' distributed,
ie. on scales comparable to 
the visible extent of normal galaxies. 
We show however that diffuse dust
is more effective at obscuring background sources.
Galaxy clusters are a likely location for ``large-scale'' diffusely
distributed dust, and its 
effect on the counts of 
background sources is explored.
We also explore the implications of a hypothesised diffuse 
intergalactic dust component uniformly distributed to high redshift
with comoving mass density equal to that 
associated with local galaxies.
In this case, we predict a deficit in background sources about three times greater than
that found in previous studies. 
\end{abstract}

{\bf Keywords:}
dust, extinction --- ISM: general --- intergalactic medium 

\bigskip

\section{Introduction}

There are a number of studies claiming that 
dust in foreground galaxies has a substantial effect
on the colours and counts of optically-selected quasars
(Ostriker \& Heisler 1984; Heisler \& Ostriker
1988; Fall \& Pei 1992 and Wright 1990).
It is estimated that at least
50\% of bright quasars at redshifts $z\simgt3$ may be obscured by foreground 
galactic dust
and hence missing from optical samples. 
These studies
assumed that dust was confined
only within the visible extent of normal massive galaxies.
However, distant populations such as faint field galaxies and quasars
may also be observed through foreground diffuse dust distributions.
Such distributions 
may be associated with galaxy clusters and extended galactic haloes.
 
A truly diffuse, intergalactic dust component is ruled out based
on the counts of quasars and reddening as a function of redshift
(eg. Rudnicki 1986; Ostriker \& Heisler 1984).
Such observations indicate that if a significant amount of dust
exists, it must be patchy and diffuse with relatively low
optical depth so that quasars will appear reddened without 
being removed from flux-limited samples.

Galaxy clusters provide a likely location for `large-scale' diffusely
distributed dust.
Indirect evidence is provided 
by several studies reporting large deficits of distant quasars
or clusters of galaxies behind nearby clusters (Boyle et al. 1988; 
Romani \& Maoz 1992
and references therein).
These studies proposed that extinction by intracluster dust was the
major cause.
Additional evidence for diffuse dust distributions is
provided by observations of massive local galaxies
where in a few cases, dust haloes extending to scales $\simgt50$kpc
have been confirmed 
(Zaritsky 1994 and Peletier et al. 1995).
 
Does uniformly distributed dust
really exist in the intergalactic medium (IGM)? 
Galactic winds associated with prodigious star formation 
at early epochs may have provided a likely source of  
metal enrichment and hence dust 
for the IGM (eg. Nath \& Trentham 1997).
Observations of metal lines in Ly-$\alpha$ absorption systems of
low column density ($N_{\rm HI}\simlt10^{15}{\rm cm^{-2}}$)
indeed suggest that the IGM was enriched to about
$Z\sim0.01 Z_{\odot}$ by redshift $z\sim3$ (Womble et al. 1996;
Songaila \& Cowie 1996). 
A source of diffuse dust may also have been provided 
by an early generation of pre-galactic stars 
(ie. population III stars) associated with
the formation of galactic haloes (McDowell 1986).
Studies have shown that possible reddening from 
uniformly distributed IGM dust 
is limited by observations of radio-selected quasars. 
Since radio-selected quasars should have 
no bias against reddening by dust, 
such a 
component must be of sufficiently low optical depth to avoid producing 
a large fraction of `reddened' sources at high redshift 
(see Webster et al. 1995; Masci 1997).
 
In this paper, we show that a given quantity
of dust has a much greater effect on the 
background universe when diffusely distributed.  
We shall investigate the effects of diffuse dust
from, first, possible distributions on
galaxy cluster scales and second,
from a hypothesised uniformly distributed component in the IGM.
 
This paper is organised as follows: 
in the next section, we explore the 
dependence of background source counts 
observed through a given mass of dust
on its spatial extent. 
In Section~\ref{spatclust}, we investigate 
the spatial distribution of dust optical depth 
through galaxy clusters and its effect on the counts 
and colours of background sources. 
Section~\ref{didu} explores the 
consequences if all dust in the local universe were assumed
uniformly distributed in the IGM.
Further implications are discussed in Section~\ref{disdc} and all results are
summarised in Section~\ref{conctwo}.
All calculations use a Friedmann cosmology with
$q_{0}=0.5$ and Hubble parameter $h_{50}=1$ where 
$H_{0}=50h_{50}\, \rm km\,s^{-1}\,Mpc^{-1}$.

\section{Compact versus Diffuse Dust Distributions}
\label{scale}

In this section, we explore 
the dependence of 
obscuration of background sources on the spatial
distribution of a given mass of dust. 
For simplicity, we assume the dust is associated with
a cylindrical face-on disk with 
uniform dust mass density. 
We quantify the amount of obscuration
by investigating the number of background sources 
behind our absorber that are 
missed from an optical flux-limited sample.
 
The fraction of sources missing to some luminosity $L$
relative to the case where there is  
no dust extinction is simply $1-N_{obs}(>L)/N_{true}(>L)$, where
$N_{obs}(>L)$
represents the observed number of sources in the presence of dust.
For a uniform dust optical depth $\tau$, $N_{obs}(>L)
\equiv N_{true}(>e^{\tau}L)$.
For simplicity, we assume that background sources are described by a 
cumulative luminosity function that follows a power-law: 
$\Phi\propto L^{-\beta}$, where $\beta$ is the slope.
This form for the luminosity function is often observed for
`luminous' ($L>L_{\ast}$) galaxies and quasars, which
dominate high redshift 
($z\simgt0.5$) populations in flux-limited samples.
With this assumption, the fraction of background sources missing
over a given area when viewed through our dusty absorber with 
uniform optical depth $\tau$ is given by
\begin{equation}
\label{fmisu}
f_{miss}(\tau)\,=1\,-\,e^{-\beta\tau}.
\end{equation}
If the `true' number of background sources per unit solid angle is 
$n_{true}$, then the total number of background sources 
lost from a flux limited sample within 
the projected radius $R$ of our absorber can be written:
\begin{equation}
\label{Nlost}
N_{lost}(\tau)\,=\,n_{true}\frac{\pi R^{2}}{D^{2}}(1\,-\,e^{-\beta\tau}),
\end{equation}
where $D$ is the distance of the absorber from us.
 
To investigate the dependence
of background source counts on the spatial dust distribution, 
we need to first 
determine the dependence of $\tau$ in equation (\ref{Nlost}) 
on the spatial extent $R$
for a {\it fixed} mass of dust $M_{d}$.
This can be determined from the individual properties of grains as follows.
The extinction optical depth at a wavelength $\lambda$ through a
slab of dust composed of grains with uniform radius $a$ is defined as
\begin{equation}
\label{tl}
\tau_{\lambda}\,=\,Q_{ext}(\lambda,a)\pi a^{2}n_{d}l_{d},
\end{equation}
where $Q_{ext}$ is the extinction efficiency which depends on the grain
size and dielectric properties, $n_{d}$ is the number density of grains
and $l_{d}$ is the length of the dust column along the line-of-sight.
Assuming our cylindrical absorber (whose axis lies along the line-of-sight) 
has a uniform dust mass density: 
$\rho_{dust}=M_{d}/\pi R^{2}l_{d}$, where $R$ is
its cross-sectional radius, 
we can write,
$n_{d}=\rho_{dust}/\frac{4}{3}\pi a^{3}\rho_{d}$,
where $\rho_{d}$ is the mass density of an individual grain.
We use the extinction efficiency $Q_{ext}$ in the $V$-band as
parameterised by Goudfrooij et al. (1994) for a graphite and silicate
mixture (of equal abundances) 
with mean grain size $a\sim0.1\mu$m characteristic of the
galactic ISM.
The value used is $Q_{ext}(V,0.1\mu m)=1.4$. 
We use a galactic extinction curve to convert
to a $B$-band extinction measure, where typically
$\tau_{B}\sim1.3\tau_{V}$ (eg. Pei 1992).
Combining these quantities,
we find that the $B$-band optical depth, 
$\tau_{B}$,
through our 
model absorber can be written in terms of its dust mass and 
cross-sectional radius as follows:
\begin{equation}
\label{to}
\tau_{B}\,\simeq\,1\left(\frac{M_{d}}{10^{8}M_{\odot}}\right)
\left(\frac{R}{20{\rm kpc}}\right)^{-2}
\left(\frac{a}{0.1\mu{\rm m}}\right)^{-1}
\left(\frac{\rho_{d}}{2{\rm gm}\,{\rm cm}^{-3}}\right)^{-1},
\end{equation}
where we have scaled to a dust mass and radius typical of local 
massive spirals and ellipticals (eg. Zaritsky 1994).
This measure is consistent with mean optical depths
derived by other means (eg. Giovanelli et al. 1994 and references therein). 

From equation (\ref{to}), we see that the dust optical depth through
our model absorber for a {\it fixed} dust mass varies in terms of 
its cross-sectional radius $R$ as 
$\tau\propto 1/R^{2}$. 
For the nominal dust parameters in equation (\ref{to}), 
the number of sources missed behind our model absorber (equation \ref{Nlost}) 
can be written
\begin{equation}
\label{NlostR}
N_{lost}(<R)\,=\,N_{true}(<20{\rm kpc})\,R_{20}^{2}(1-e^{-\beta R_{20}^{-2}}),
\end{equation}
where $R_{20}=R/20{\rm kpc}$ and 
\begin{equation}
N_{true}(<20{\rm kpc})\,\equiv\,n_{true}\frac{\pi(20{\rm kpc})^{2}}{D^{2}}
\label{Ntrues}
\end{equation}
is the `true' number of background sources falling
within the projected scale radius $R=20$kpc.
 
From the functional forms of equations (\ref{Nlost}) and (\ref{NlostR}), there
are two limiting cases:
\begin{enumerate}
\item
For optical depths $\tau\gg1/\beta$, the factor 
$1-e^{-\beta\tau}$ in equation (\ref{Nlost}) is of order unity. 
This corresponds to values of $R$ such that $0<R_{20}\ll\beta^{1/2}$
for the nominal parameters in equation (\ref{to}).
For values of $R$ in this range, we have $N_{lost}\propto R^{2}$ and the 
obscuration of background sources will depend most strongly on $R$.
 
\item
For $\tau\ll 1/\beta$ or equivalently $R_{20}\gg\beta^{1/2}$,
$N_{lost}$ will approach a constant limiting value,
independent of the dust extent $R$. From equation (\ref{NlostR}), this
limiting value can be shown to be 
$N_{lost}(<R)\,=\,N_{true}(<20{\rm kpc})\beta$.
\end{enumerate}
 
\begin{figure}
\vspace{-2.5in}
\plotonesmall{0.9}{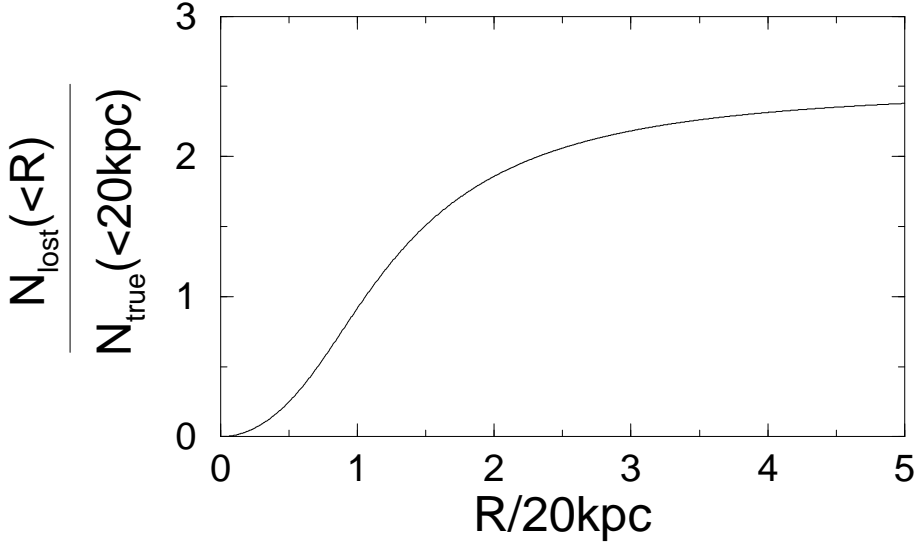}
\vspace{-2.7in}
\caption{The number of sources missing from an optical flux-limited sample
behind a face-on dusty disk
with {\it fixed}
mass $M_{dust}=10^{8}M_{\odot}$ as a function of its radial extent $R$.
This number is normalised against the `true' number of background sources
(in the absence of the absorber) that fall within the solid angle
$\pi(20{\rm kpc})^{2}/D^{2}$ subtended by our nominal
radius of 20kpc at distance $D$. See equation (\ref{NlostR}.)}
\label{NlostRgal}
\end{figure}

As a simple illustration, we show in Figure \ref{NlostRgal} the
dependence of the number of background sources 
missing behind our model dust absorber on
$R$, for a fixed dust mass of
$10^{8}M_{\odot}$ as defined by equation (\ref{NlostR}).
We assumed a cumulative luminosity function slope of $\beta=2.5$,
typical of that for luminous galaxies and quasars. 
From the above discussion, we see that when 
$R_{20}\simgt\beta^{1/2}$, ie. $R\simgt30{\rm kpc}$
(or when $\tau\simlt 1/\beta=0.4$), 
the obscuration
will start to 
approach its maximum value and remain approximately constant as 
$R\rightarrow\infty$. 

We conclude
that when dust becomes diffuse and extended on a scale such that
the mean optical depth $\tau$ through the distribution
satisfies $\tau<1/\beta$, where $\beta$ is the cumulative luminosity
function slope of background sources, obscuration 
will start to be important and is maximised for $\tau\ll1/\beta$.
The characteristic spatial scale at which this occurs  
will depend on the dust mass through equation (\ref{to}).
For the typical grain values in equation (\ref{to}),
this characteristic radius
is given by
\begin{equation}
R\,\simeq\,31\left(\frac{\beta}{2.5}\right)^{1/2} 
\left(\frac{M_{d}}{10^{8}M_{\odot}}
\right)^{1/2}{\rm kpc}. 
\label{radiuscale}
\end{equation}
The simple model in Figure \ref{NlostRgal} shows that the
obscuration of background sources due 
to a normal foreground galaxy will be most effective if dust
is distributed over a region a few times its optical radius.
This prediction may be difficult to confirm observationally
due to possible contamination from light in the galactic absorber.
In the following
sections, we explore
two examples of possible `large-scale' diffuse 
dust distributions that can be explored 
observationally.

\section{Diffuse Dust in Galaxy Clusters}
\label{spatclust}

A number of studies
have attributed the existence of large deficits of 
background sources behind nearby galaxy clusters as 
due to extinction by dust. 
Bogart \& Wagner (1973) found that
distant rich Abell clusters were anticorrelated on the sky with nearby ones.
They argued for a mean extinction of $A_{V}\approx0.4$ mag
extending to $\sim2.5$ times the optical radii of the nearby clusters.
Boyle et al. (1988)
however claimed a 
$\sim30\%$ deficit of background quasars within $4'$ of clusters
consisting of tens of galaxies.
These authors
attribute this to an extinction $A_{V}\approx0.15$ mag, and
deduce a dust mass of $\sim10^{10}M_{\odot}$ within 0.5Mpc 
of the clusters. 
Romani \& Maoz (1992) found that optically-selected quasars
from the V\'{e}ron-Cetty \& V\'{e}ron (1989) catalogue avoid
rich foreground Abell clusters.
They also found deficits of $\sim30\%$
out to radii
$\sim5'$ from the clusters, and postulate a
mean extinction, $A_{V}\approx0.4$ mag.
 
The numbers of background sources behind clusters is also expected to
be modified by 
gravitational lensing (GL) by the cluster potential.
Depending on the intrinsic luminosity function of the background population,
and the limiting magnitude to which the sources are detected, GL 
can cause either an enhancement or a deficit in the number of 
background sources. 
The GL effect has been used to explain various 
reports of overdensities of both optically and radio-selected
quasars behind foreground clusters
(Bartelmann \& Schneider 1993; Bartelmann et al. 1994;
Rodrigues-Williams \& Hogan 1994;
Seitz \& Schneider 1995).  
The reported overdensities for {\it optically-selected} QSOs 
are contrary to the studies above where 
anticorrelations with foreground 
clusters are found.
These overdensities however are claimed on angular scales
$\sim10'-30'$ from the cluster centers, considerably
larger than scales on which most of the underdensities have been claimed,
which are of order a few arcminutes. 
One interpretation is that
dust obscuration bias may be greater towards cluster centers due 
to the presence of greater quantities of dust.
On the other hand, the reported anticorrelations on small
angular scales can perhaps be explained by the optically 
crowded fields, where QSO identification may be difficult.
At present, the effects of clusters on background source counts still
remains controversial. 
 
More direct evidence for the existence of intracluster dust
was provided by Hu et al. (1985) and 
Hu (1992) who compared the Ly-$\alpha$ flux from
emission line systems in ``cooling flow'' clusters with
Balmer line fluxes at optical wavelengths.  
Extinctions of
$A_{V}\sim0.2-1$ mag towards the cluster centers
were found, in good agreement with estimates
from the quasar deficits above.
Intracluster dust has been predicted to give rise to
detectable diffuse IR emission (eg. Dwek et al. 1990).
The most extensive search was conducted by Wise et al. (1993), who
claimed to have detected excess diffuse 60-100$\mu$m emission at the 2$\sigma$
level from a number of rich Abell clusters.
They derived dust temperatures in the range 24-34 K and
dust masses $\sim10^{10}M_{\odot}$ within radii of $\sim1$Mpc.
Recently,
Allen (1995) detected strong X-ray absorption and optical 
reddening in ellipticals situated at the centers of rich cooling flow
clusters, providing strong evidence for dust. 
These studies indicate that intracluster dust is
certainly present, however, 
the magnitude of its effect on producing background source
deficits remains a controversial issue. 
 
In this section, we
give some predictions that
may be used to further constrain cluster dust properties, or
help determine the dominant mechanism (ie. GL or extinction) by which clusters
affect background observations. 

\subsection{Spatial Distribution of Cluster Dust} 

X-ray spectral measurements show the presence of hot, metal enriched
gas in rich galaxy clusters with $\sim0.5-1$ solar metallicity.
This gas is believed to be both
of galactic and primordial
origin (ie. pre-existing IGM gas), with the bulk of metals being 
ejected from cluster galaxies (see Sarazin 1986 for a review).
Ejection from galaxies may occur abruptly
through collisions between the cluster galaxies, `sudden' ram-pressure
ablation, or through continuous ram-pressure stripping by
intracluster gas (eg. Takeda et al. 1984). 
The lack of significant amounts of dust (relative to what should have
been produced by stellar evolution) and interstellar gas in cluster
ellipticals provides evidence for a mass loss process.
On the other hand, in ellipticals that avoid dense cluster
environments, significant quantities 
of neutral hydrogen, molecular gas and dust 
have been detected (eg. Lees et al. 1991).
 
If the dust-to-gas ratio of intracluster gas 
in rich clusters were similar to that of the Milky Way, then a radial
gas column density of typically $\simgt10^{22}{\rm cm}^{-2}$
with metallicity $Z=0.5Z_{\odot}$ would produce an extinction 
$A_{V}\simgt4$ mag.
This however is much greater than the value observed. 
The likely reason for the deficiency of dust in the intracluster medium is its
destruction by thermal sputtering in the hot gas, a process which operates on
timescales $t_{sputt}\simlt10^{8}n_{-3}a_{0.1}$ yr, where
$n_{-3}=n_{H}/10^{-3}{\rm cm}^{-3}$ is the gas density and
$a_{0.1}=a/0.1\mu{\rm m}$ the grain radius (Draine \& Salpeter 1979). 
Dust injection timescales from galaxies
is typically of order a Hubble time 
(eg. Takeda et al. 1984) and 
hence, grains are effectively destroyed, with only the most recently
injected still surviving and providing possibly some measurable 
extinction.
 
The spatial distribution in dust mass density remains a 
major uncertainty. A number of authors have shown that
under a steady state of continuous injection from
cluster galaxies, destruction by thermal sputtering
at a constant rate, and assuming instantaneous mixing with the hot gas,
the resulting mass density in dust will be of order 
\begin{equation}
\label{rhodust}
\rho_{dust}\,\sim\,
10^{-31}\left(\frac{a}{0.1\mu{\rm m}}\right)
\left(\frac{Z_{d}}{0.01}\right)h_{50}\,\,{\rm gm}\,{\rm cm}^{-3},
\end{equation}
(eg. Dwek et al. 1990)
where $Z_{d}$ is the injected dust-to-gas mass ratio, assumed to be
equal to the mean value of the galactic ISM, $Z_{d}\simeq0.01$ (Pei 1992). 
According to this simple model,
the dust mass density is independent of gas density and position in the
cluster.
If we relax the assumption of instantaneous mixing of dust with
the hot gas however, so that the spatial distribution of gas is
different from that of the injected dust, the radial distribution
of dust can significantly differ from uniformity throughout a cluster.
Such a non-uniform spatial dust distribution
may be found in clusters exhibiting cooling flows. 
If, as suggested by Fabian et al. (1991), most of the cooled gas remains
cold and becomes molecular in cluster cores, then
a relatively large amount of dust may also form, resulting in a
dust distribution which peaks within the central regions. 
 
We explore the radial dependence of extinction
optical depth through a cluster,
and the expected deficit in background sources
by assuming that dust is diffusely distributed and follows a
spatial density distribution:
\begin{equation}
\label{rhodR}
\rho_{dust}(R)\,=\,\rho(0)\left[1 - \left(\frac{R}{R_c}\right)^2\right]^{-n},
\end{equation}
where $R_c$ is a characteristic radius which we fix and 
$n$ is our free parameter. 
Equation (\ref{rhodR}) with $n=3/2$ is the usual King profile which
with $R_c\simeq0.25$Mpc, represents a good approximation to 
the galaxy distribution in clusters.
Thus for simplicity, we keep $R_c$ fixed at $R_c=0.25$Mpc and vary $n$.
To bracket the range of possibilities in the distribution of 
intracluster dust, we consider the range $0\,<\,n\,<\,3/2$. 
$n=0$ corresponds to the simple case where 
$\rho_{dust}(R)=\rho(0)=$constant, which may describe a situation
where injection of dust is balanced by its destruction by hot gas 
as discussed above.
The value $n=3/2$ assumes that dust
follows the galaxy distribution. This profile may arise 
if grain destruction by a similar distribution of 
hot gas were entirely absent. 

\subsection{Spatial Distribution of Dust Optical Depth 
and Background Source Deficits}
 
To model the spatial distribution of optical depth,
we assume that intracluster dust is distributed within a 
sphere of radius $R_{max}$. 
The central dust mass density $\rho(0)$ in equation (\ref{rhodR})
is fixed by assuming
values for the total dust mass $M_{dust}$ and $R_{max}$ such that 
\begin{equation}
\label{Mdust2}
M_{dust}=\int^{R_{max}}_0{4\pi R^{2}\rho(R)\,dR}.
\end{equation}
We assume a cluster dust radius of $R_{max}=1$Mpc, which represents
a radius containing $\simgt90\%$ of the virial mass 
of a typical dense cluster characterised by 
galactic velocity dispersion $\sigma_{V}\sim800{\rm km s}^{-1}$ 
(Sarazin 1986 and references therein).   
We assume 
that the total dust mass within $R_{max}=1$Mpc 
is $M_{dust}=10^{10}M_{\odot}$. This value is
consistent with that derived from extinction measures by 
Hu, Cowie \& Wang (1985), IR emission detections by Wise et al. (1993) and
theoretical estimates of 
the mean intracluster dust density as given by equation (\ref{rhodust}). 
 
Using equation (\ref{tl}), the $B$-band optical depth through our
spherical intracluster dust distribution at some projected distance
$r$ from its center can be written: 
\begin{equation}
\label{tint}
\tau_{B}(r)\,=\,\frac{3Q_{ext}}{4a\rho_{d}}\int^{2(R_{max}^{2}-r^{2})^{1/2}}
_{0}{\rho_{dust}\left([r^{2}+R'^{2}]^{1/2}\right)\,dR'}, 
\end{equation}
where $\rho_{dust}(R)$ is our assumed radial density distribution 
(equation \ref{rhodR}) and $\rho_{d}$ the mass density of an individual dust grain. 
For a uniform dust density ($\rho_{dust}(R)=\rho_{dust}(0)=$constant),
and our assumed values of $R_{max}$ and $M_{dust}$ given above,
the radial dependence in dust optical depth can be written:
\begin{equation}  
\label{toclr}
\tau_{B}(r)\,=\,\tau_{B}(0)\left[1-
\left(\frac{r}{R_{max}}\right)^{2}\right]^{1/2}, 
\end{equation}
where $\tau_{B}(0)$ is the optical depth through the center of our cluster,
which with 
grain properties characteristic of the galactic ISM, will scale as
\begin{equation}
\label{tocl}
\tau_{B}(0)\,\simeq\,0.06\left(\frac{M_{dust}}{10^{10}M_{\odot}}\right)
\left(\frac{R_{max}}{{\rm Mpc}}\right)^{-2}
\left(\frac{a}{0.1\mu{\rm m}}\right)^{-1}
\left(\frac{\rho_{d}}{2{\rm gm}\,{\rm cm}^{-3}}\right)^{-1}.
\end{equation}
This value is about a factor three times lower than estimates of the
mean extinction derived from the deficit of QSOs
behind foreground clusters (eg. Boyle et al. 1988), and that implied by the
Balmer decrements of Hu et al. (1985).
For a fixed dust mass of $10^{10}M_{\odot}$ however, we can achieve
larger values for the central optical depth by steepening the 
radial dust-density distribution profile, determined by the slope $n$
in equation (\ref{rhodR}). 
 
\begin{figure}
\vspace{-3in}
\plotonesmall{0.9}{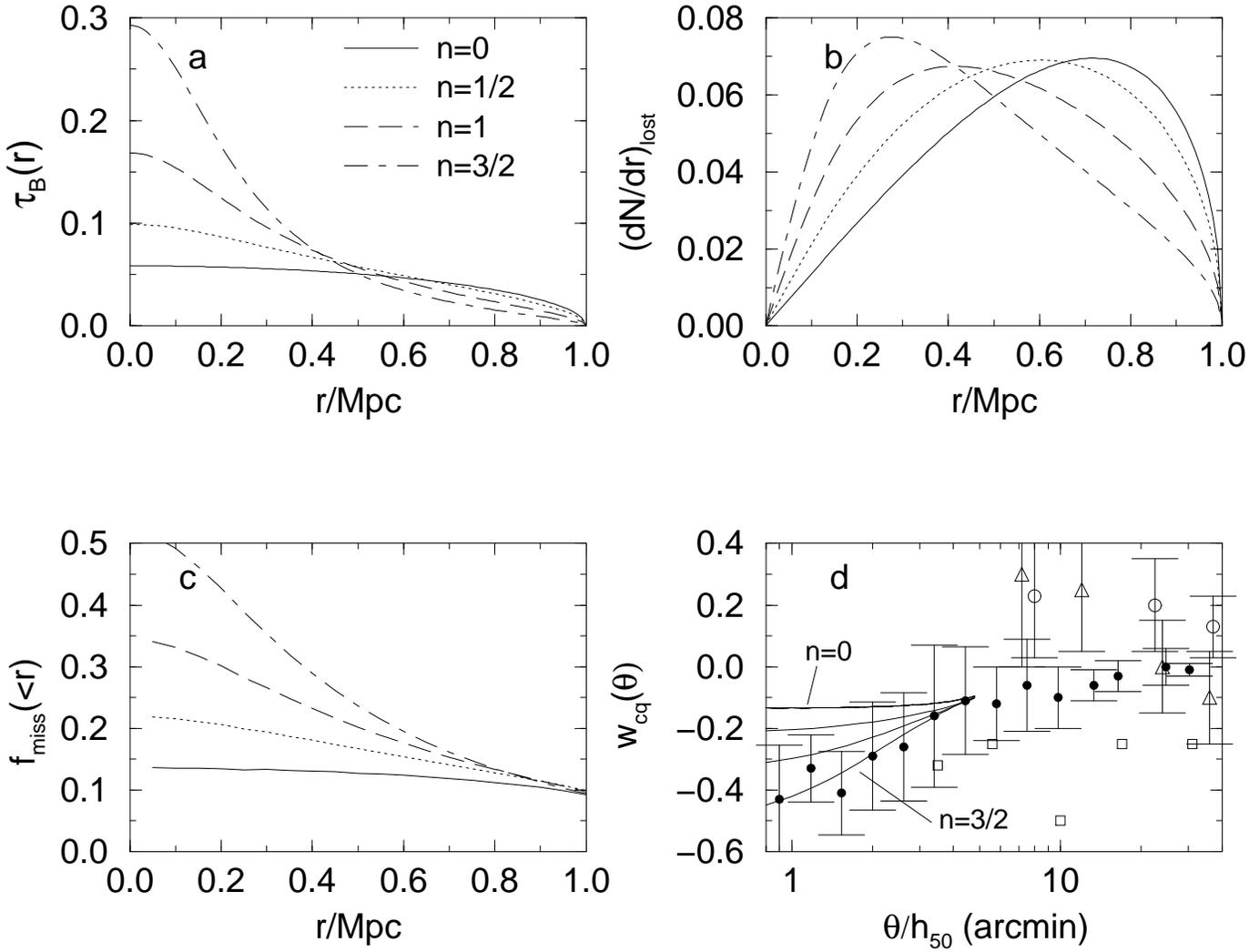}
\vspace{-1in}
\caption{(a) Optical depth as a function of projected cluster radius for
various dust density distributions as parameterised by equation (\ref{rhodR}).
(b) Differential number of background sources lost from a flux-limited 
sample (arbitrary scale). 
(c) Total fraction of background sources missing to some $r$ 
and (d) Cluster-QSO two-point angular correlation function,
filled circles: Boyle et al. (1988), squares: Romain \& Maoz (1992),
open circles: Rodrigues-Williams \& Hogan (1994), triangles: 
Rodrigues-Williams \& Hawkins (1995).} 
\label{clusters}
\end{figure}

Figure \ref{clusters}a shows the optical depth
as a function of projected cluster radius
for the cases $n=0$, 0.5, 1, and 1.5.
The case $n=0.5$ approximately corresponds to
the model of Dwek et al. (1990), which included effects of mild
sputtering by hot gas in order to fit for the observed IR emission
from the Coma cluster. 
As shown, the case $n=0$ ($\rho_{dust}(R)=$constant) predicts that 
the dust optical depth should be almost 
independent of projected radius $r$.
Within all projected radii, the optical depths predicted by 
our diffuse dust model lie in the range $0<\tau_{B}<0.3$.
Turning back to the discussion of Section~\ref{scale} where we show that 
background obscuration by diffuse dust reaches its maximum 
for $\tau_{B}<1/\beta$, these optical depths satisfy this
condition for $\beta\simlt2.5$, typical of luminous 
background galaxies and quasars. 

We now explore the effects of these models 
on background source counts as a function of projected cluster radius.
We first give an estimate of the projected radius 
at which the numbers of background sources lost from a flux-limited
sample is expected to be a maximum.
This is determined by investigating the dependence in the differential 
number of
sources missing, $dN_{lost}$, within an interval ($r,r+dr$) as a 
function of projected radius $r$. 
From equation (\ref{Nlost}), this 
differential number will scale as 
\begin{equation}
\label{dNlost}
dN_{lost}(r)\,\propto\,r\,dr\left[1 - e^{-\beta\tau(r)}\right],
\end{equation} 
where $\tau(r)$ is given by equation (\ref{tint}).
Figure \ref{clusters}b plots $dN_{lost}/dr$ as a function of $r$ for 
our various models, where we have assumed $\beta=2.5$.
Thus from observations, 
an identification of the projected radius at which the 
background source deficit peaks can be used to  
constrain the spatial distribution of intracluster dust. 
 
The cumulative fraction of background 
sources missing within a projected cluster
radius is given by 
\begin{equation}
\label{Nlostltr}
f_{miss}(<r)\,=\,\frac{N_{lost}(<r)}{N_{true}(<r)}\,=\,
\frac{1}{\pi r^{2}}\int^{r}_{0}{2\pi r'\left[1-e^{-\beta\tau(r')}\right]\,dr'}. 
\end{equation}
This fraction is shown in Figure \ref{clusters}c. 
As expected, the $n=3/2$ model which contains the largest amount of dust
within the inner few hundred kiloparsecs predicts the strongest
trend with $r$, while the opposite is predicted if the dust density 
were completely uniform. 
These predictions can be compared with a number of existing studies of the
observed two-point angular correlation function between clusters and 
optically-selected QSOs. 
This function is usually defined as 
\begin{equation}
\label{wtheta} 
w_{cq}(\theta)\,=\,\frac{\langle N_{obs}(<\theta)\rangle} 
{\langle N_{ran}(<\theta)\rangle}\,-\,1, 
\end{equation}
where $\langle N_{obs}(<\theta)\rangle$ is the average number
of observed cluster-QSO pairs within an angular radius $\theta$ and
$\langle N_{ran}(<\theta)\rangle$ is that expected in a random
distribution.
For our purposes, $\langle N_{ran}(<\theta)\rangle$ can be
replaced by the ``true'' number of cluster-QSO pairs expected in the
absence of dust,
and hence, we can re-write equation (\ref{wtheta}) as 
\begin{equation}
\label{wthetalost}
w_{cq}(\theta)\,\equiv\,-\frac{\langle N_{lost}(<\theta)\rangle}
{\langle N_{true}(<\theta)\rangle}\,=\,-f_{miss}(<\theta). 
\end{equation}
 
We compare our models with a number of studies of $w_{cq}(\theta)$
for optically selected QSOs in Figure \ref{clusters}d.
These studies considerably differ from each other in the selection
of the QSO and cluster samples, and as seen, both 
anticorrelations and correlations on different angular scales 
are found.
The former have been interpreted in terms of extinction by intracluster
dust, while the latter with the GL phenomenon.
In most cases, the reported overdensities are too large to be 
consistent with GL models given our current knowledge
of cluster masses and QSO distributions.
 
It is interesting to note that the studies which have reached the
smallest angular scales ($\simlt5'$) are also those in which anticorrelations
between QSOs and foreground clusters have been reported. 
This can be understood in terms of a larger dust concentration
and hence extinction towards cluster centers.
These studies however may not be free of selection effects,
such as in the detection of QSOs from the visual inspection of
objective prism plates. 
From a cross-correlation analysis of galactic stars with their cluster
sample however, 
Boyle et al. (1988) found that such selection effects are minimal.
 
The maximum dust radial extent assumed in our models, 
$R_{max}=1$Mpc, corresponds to angular scales $\sim5'$ at the mean 
redshift of the clusters ($\langle z_{c}\rangle\sim0.15$) used in
these studies. 
Thus, as shown in Figure \ref{clusters}d, 
our model predictions only extend to $\sim5'$.
As shown in this figure, the $n=3/2$ model which corresponds to the
case where the dust density is assumed to follow the galaxy distribution,
provides the best fit to the Boyle et al. (1988) data. 
We must note that this is the only existing study performed to 
angular scales $\sim1'$ with which we can compare our models. 
Further studies to such scales are necessary to 
confirm the Boyle et al. result, and/or 
provide a handle on any selection effects.  

\subsection{Summary}
 
To summarise, we have shown that for a plausible value of the
dust mass in a typical rich galaxy cluster, 
obscuration of background sources will be most
effective if dust is diffusely distributed on scales $\sim1$Mpc.
This conclusion is based on our predicted optical depths
($\tau_{B}<0.3$) satisfying our condition for 
`maximum' obscuration: $\tau_{B}<1/\beta$ 
(see Section~\ref{scale}), where typically
$\beta\simlt2.5$ for luminous background galaxies and QSOs. 
 
We have explored the spatial distribution in dust optical depth
and background source deficits expected through a 
typical rich cluster by assuming different 
radial dust density profiles. 
These predictions can be used to constrain cluster dust properties.
A dust density distribution with $n=3/2$ (equation \ref{rhodR}) appears
to best satisfy the `small scale' cluster-QSO angular correlation study
of Boyle et al. (1988). 
 
\section{Diffuse Intergalactic Dust?}
\label{didu}
 
There have been a number of studies claiming that
the bulk of metals in the local universe had already formed by  
$z\sim1$ (eg. Lilly \& Cowie 1987; White \& Frenk 1991; Pei \& Fall 1995). 
Similarly, models of dust evolution in the galaxy show that the
bulk of its dust content was formed in the first few billion years
(Wang 1991).
These studies suggest that 
the global star formation rate peaked at epochs $z\simgt2$ 
when the bulk of 
galaxies were believed to have formed.
Supernova-driven winds at early epochs
may thus have provided 
an effective mechanism by which
chemically enriched material and dust were dispersed into the IGM.
As modelled by Babul \& Rees (1992), such a mechanism
is postulated to be crucial in the evolution of the `faint blue'
galaxy population observed to magnitudes $B\sim28$.
Nath \& Trentham (1997) also show that this mechanism could explain 
the recent detection of
metallicities $Z\sim0.01Z_{\odot}$
in low density Ly-$\alpha$ absorption systems at $z\sim3$. 
Another source of diffuse IGM dust may have been provided by 
an epoch of population III star formation associated with the
formation of galactic haloes
(eg. McDowell 1986).

What are the effects expected on background sources if all dust formed
to the present day was
completely uniform and diffuse throughout the IGM?
In this section, we show that 
such a component will have a low optical depth and have an
insignificant effect 
on the colours of background sources, but will be high enough
to significantly bias their number counts 
in the optical.
 
\subsection{Comoving Dust Mass Density}
 
To explore the effects of a diffuse intergalactic dust component, we need to 
assume a value for the mean mass density in dust in the local universe.
This density must not exceed the total mass density in heavy metals at the
present epoch. 
An upper bound for the local mass density in metals (hence dust)
can be derived from the assumption that the {\it mean} metallicity of the
local universe is typically: $Z\sim\Omega_{metals}/\Omega_{gas}\sim0.01$ 
(ie. the ratio of elements heavier than helium to total gas mass), 
as found from galactic
chemical evolution models (eg. Tinsley 1976) and abundance observations
(Grevesse \& Anders 1991). 
Combining this with the upper bound in the baryon density predicted from
big-bang nucleosynthesis (Olive et al. 1990) where 
$\Omega_{gas}\simlt\Omega_{baryon}<0.06h_{50}^{-2}$, it is apparent that
\begin{equation}
\label{upperlim}
\Omega_{metals}(z=0)\,<\,6\times10^{-4}h_{50}^{-2}.
\end{equation}
 
Let us now compute the total mass density in dust used in
previous studies that modelled the effects of dust in 
individual galaxies on background quasars.
Both Heisler \& Ostriker (1988) and Fall \& Pei (1993) modelled
these effects by assuming that dust in each galaxy
was distributed as an exponential disk with scale radius 
$r_{0}\simeq30$kpc and central face-on optical depth, $\tau_{B}=0.5$. 
The comoving mean mass density in dust (relative to the critical density) 
in these studies,
given a comoving galaxy number density
$n_{0}=0.002h_{50}^{3}{\rm Mpc}^{-3}$, can be shown to be
\begin{equation}
\label{HOdust}
\Omega_{dust0}\,\simeq\,7.3\times10^{-6}h_{50}
\left(\frac{n_{0}}{0.002{\rm Mpc}^{-3}}\right) 
\left(\frac{r_{0}}{30{\rm kpc}}\right)^{2}\left(\frac{\tau_{B}}{0.5}\right) 
\end{equation}
(see Masci 1997).
This is consistent with the constraint in equation (\ref{upperlim}).
Thus, as a working measure, we assume the comoving mass
density defined by equation (\ref{HOdust}) in the calculation that follows. 
 
\subsection{Obscuration by Diffuse Intergalactic Dust}
 
If the dust mass density given by equation (\ref{HOdust}) is assumed 
uniformly distributed and constant 
on comoving scales to some redshift,
the $B$-band optical depth through a dust sheet of width d$l$ at redshift $z$
in an observer's frame can be written (see equation \ref{tl})
\begin{equation}
\label{dtau}
d\tau_{B}\,\simeq\,7.2\times10^{-6}h_{50} 
\left(\frac{\Omega_{dust0}}{7.3\times10^{-6}}\right)
\left(\frac{a}{0.1\mu{\rm m}}\right)^{-1}
\left(\frac{\rho_{d}}{2{\rm gm}\,{\rm cm}^{-3}}\right)^{-1} 
\left(\frac{{\rm d}l}{{\rm Mpc}}\right)(1+z),
\end{equation}
where for simplicity, 
we have assumed dust properties characteristic of the galactic ISM.
The factor $(1+z)$ is due to our assumption of a $1/\lambda$
dependence for the dust extinction law. This arises
from the fact that light received in the $B$-band 
corresponds to light of wavelength $\lambda_{B}/(1+z)$ at redshift $z$,
which consequently suffers greater extinction.
Although a $1/\lambda$ law is not fully representative
of that observed in the galactic ISM which includes the strong 
2200\AA~ feature, this is a good on average ralation for the dust
laws in many external galaxies (eg. Jansen et al. 1994). Such an assumption
greatly simplifies the redshift dependence of extinction in an
observer's frame. 
With $dl/dz=6000h_{50}^{-1}(1+z)^{-5/2}$Mpc (for a $q_{0}=0.5$ and 
$\Lambda=0$ cosmology), the total mean optical depth
to some redshift in an observer's $B$-band will scale as
\begin{equation}
\label{tautotone} 
\tau_{B}(z)\,\simeq\,0.1\left(\frac{\Omega_{dust0}}{7.3\times10^{-6}}\right)
\left(\frac{a}{0.1\mu{\rm m}}\right)^{-1}
\left(\frac{\rho_{d}}{2{\rm gm}\,{\rm cm}^{-3}}\right)^{-1}
\left[1-(1+z)^{-1/2}\right].
\end{equation}
This represents the total optical depth if all dust in the 
intervening galaxy model of Heisler \& Ostriker (1988) were
assumed uniformly distributed throughout the universe.
 
Assuming dust is uniformly distributed to $z=2$, the {\it observed} 
$B$-band optical
depth from equation (\ref{tautotone}) will be of order
\begin{equation}
\label{tauz2}
\tau_{B}({\rm U})(z=2)\,\simeq\,0.04.
\end{equation}
Using a galactic extinction law (Pei 1992), 
this corresponds to an extinction in $B-R$ colour of 
$E_{B-R}\sim0.02$mag.
Thus, if background faint field galaxies and QSOs are observed through
a uniform intergalactic dust distribution, their observed colours are
not expected to be significantly affected. 
We now show however that the numbers of sources missing at such redshifts 
could be significantly greater than that claimed by previous studies which 
assume all dust to be associated with
massive galaxies alone.
 
If dust to some distance $D$ covers an area of sky $A$
and hence, has covering factor $C_{d}=A/4\pi D^{2}$,
the number of
background sources lost from a flux-limited sample can be estimated from
equation (\ref{Nlost}). 
In general, the number of background sources at some redshift
lost from
an area of sky with dust covering factor $C_{d}$ will scale as: 
\begin{equation}
\label{Nlostgen}
N_{lost}(\tau,z)\,\propto\,C_{d}f_{miss}(\tau,z), 
\end{equation}
where $f_{miss}\equiv 1-e^{-\beta\tau_{B}(z)}$
is the fraction of sources missing per unit area. 
For a completely uniform dust distribution, 
$C_{d}({\rm U})=1$, and to 
redshift $z=2$, $f_{miss}\simeq10\%$ for $\beta=2.5$.  
 
If dust were confined to individual galaxies along the
line-of-sight however, their covering factor, 
assuming they follow a  
Poisson distribution is typically
$C_{d}\simeq\bar{N}_{z}\exp{(-\bar{N}_{z})}$,
where $\bar{N}_{z}$ is the mean number of absorber intersections 
to redshift $z$:
\begin{equation}
\label{Nz1}
\bar{N}_{z}\,\simeq\,0.01h_{50}^{2}
\left(\frac{n_{0}}{0.002{\rm Mpc}^{-3}}\right)
\left(\frac{r_{0}}{30{\rm kpc}}\right)^{2}\left[(1+z)^{1.5} - 1\right]
\end{equation}
(see Heisler \& Ostriker 1988).
We have scaled to the nominal parameters assumed in the
intervening galaxy model of Heisler \& Ostriker (1988) (hereafter HO). 
In this model, we find a covering
factor of only $C_{d}({\rm HO})\simeq0.04$ to $z=2$.
We can estimate the mean effective optical depth {\it observed} in the
$B$-band through an individual absorber to $z\simeq2$ in the HO model
by using the formalism of Section~\ref{scale}. 
For a {\it fixed} mass of dust, equation (\ref{to}) implies that the
product of the area (or covering factor) and optical depth of 
a dust distribution:
$\tau\times C_{d}$, is a constant, depending 
on grain properties and dust mass alone.
Using this relation,
the {\it observed} effective absorber optical depth to $z\simeq2$ 
in the HO model, $\tau_{B}({\rm HO})$, 
can be estimated by scaling from our values 
of $\tau_{B}({\rm U})$ and $C_{d}({\rm U})$ 
above for uniformly distributed dust:
\begin{equation}
\tau_{B}({\rm HO})\,\simeq\,\tau_{B}({\rm U})C_{d}({\rm U})/C_{d}({\rm HO})
\,=\,\frac{0.04\times1}{0.04}\,=1.
\label{HOopt}
\end{equation}
Using this value, the fraction of background sources
missed by obscuration from an individual absorber is `effectively' 
$f_{miss}=1-\exp(-2.5\times1)\simeq91\%$. 
Combining these results, we find using equation (\ref{Nlostgen}) that
the number of sources missing at $z\simeq2$ due to a uniform
foreground dust distribution to be greater by a factor of 
$\frac{1\times0.1}{0.04\times0.91}\,\sim\,3$
than that predicted by Heisler \& Ostriker (1988). 
 
We must note that this estimate makes no                      
allowance for possible evolution in dust content. 
Effects of foreground diffuse dust on source counts           
at $z>2$ may be significantly reduced 
if appreciable evolution has occured.
Effects of models where the dust content evolves have been explored by
Masci \& Webster (1998)
 
\subsection{Summary}
 
We conclude that the existence of a significant amount of
diffusely distributed dust (eg. with mass density 
on comoving scales of order that 
observed in local galaxies) can enhance the
number of background sources missing in optical samples. 
Due to its relatively large covering factor, diffuse dust predicts 
a reduction in optical counts at $z>2$ about three 
times greater than
that claimed by previous studies.
 
The colours of background sources
are not expected to be significantly affected.
This implies that the use of background populations 
to measure a diffuse IGM dust component will be extremely difficult. 
 
\section{Discussion}
\label{disdc}
 
In this section, we discuss some further uncertainties 
and implications regarding the existence of diffuse dust in the universe.  
 
First, the effects of intracluster dust on background sources
critically depends 
on the amount of dust present, and its spatial distribution. 
Regardless of the mechanism by which grains are
injected into the intracluster medium from galaxies,
it is possible that a 
significant fraction are destroyed in the injection
process.
Significant amounts of hot gas are also believed to exist in the
ISM of cluster ellipticals (eg. Forman et al. 1985). 
This gas
is expected to destroy grains on timescales
$\simlt10^{8}$yr (Draine \& Salpeter 1979), 
much shorter than injection timescales.
Such destruction mechanisms can thus prevent the
formation of significant quantities
of dust.
 
It is possible that
the spatial distribution of intracluster dust is not `diffuse'
and uniformly distributed, but inhomogeneous.
For example, Fabian et al. (1991) propose that if most of the cooled
gas resulting from cluster cooling flows remains cold and becomes
molecular, then this may provide suitable conditions for 
large amounts of dust to form.
A clumpy dust distribution that follows cooling flow filaments
may result, reducing the effective dust covering factor
and hence background source deficit. 
These issues need to be addressed
before attributing such deficits
to extinction by dust.
 
The existence of a smooth IGM dust component also remains
a major uncertainty.
Due to their deep gravitational potential, Margolis \& Schramm (1977) 
showed that it is unlikely for supernovae-driven winds to 
expel significant quantities 
of dust from a massive galaxy to large scales.
For low mass galaxies however (eg. dwarfs), Babul \& Rees
(1992) show that this mechanism can be effective. 
Such systems are believed to comprise a majority of the
`faint-blue population' which show an excess $\sim20-30$ times that 
predicted from non-evolving galaxy models for $B>24$ (eg. Tyson 1988). 
Simulations 
based on star-formation rates that assume yields in metallicity from
local observations, predict that the
amount of metals (and hence dust, assuming a fixed fraction of
metals condense into grains at a constant rate) produced from such a population 
will be smaller than local estimates by an order of magnitude
(eg. White \& Frenk 1991). 
If the only source of IGM dust was from these `low-mass' galaxies, 
then the total optical depth to $z\simgt2$ would
be insignificant, and effects
on the background universe would be minimal. 
 
\section{Conclusions}
\label{conctwo}
 
In this paper, we have shown that dust is more effective
at obscuring background sources
when diffuse or extended.
We find that obscuration of
background sources by a given dust distribution with
optical depth $\tau_{B}$ will be most effective 
when $\tau_{B}<1/\beta$, where $\beta$ is the 
cumulative luminosity function slope of the sources. 
 
We have explored the effects of diffuse dust from, first, galaxy clusters 
and second, from a hypothesised uniform IGM component.
By assuming different radial dust density profiles
in a typical rich cluster,
we have predicted the optical depth and background source deficit
as a function of projected cluster radius. 
These predictions can be compared with future observations 
to constrain the properties of intracluster dust.
Our predicted optical depth measures ($\tau_{B}\simlt0.3$) 
satisfy the above criterion ($\tau_{B}<1/\beta$) for background luminous
QSOs and
galaxies.  
Existing studies claiming anticorrelations in the
distribution of QSOs with foreground clusters down to scales $\sim1'$
are consistent with a dust density profile that follows
the galaxy distribution. 
 
As a further illustration,
we have explored the effects of a diffuse IGM dust component
with cosmic mass density equal to that observed 
in local galaxies. 
Assuming this density is constant on
comoving scales to $z=2$, 
we find
a deficit in background sources about three 
times greater than that predicted 
assuming dust in normal galaxies alone.

The `diffuseness' of the dust is the key parameter which we claim
determines the effectiveness of obscuration of the background universe.  
Although such dust distributions may be difficult to detect,
we must not neglect their possible presence. 
Further studies of spatial dust distributions, 
preferably via 
the counts and colours of background sources will be essential
in confirming our predictions.

\section*{Acknowledgements}

I am grateful to Paul Francis and Rachel Webster for many illuminating discussions
and suggestions.
I also acknowledge support from an Australian Postgraduate Award.

\section*{References}

\reference Allen, S. W. 1995, MNRAS, 276, 947

\reference Babul, A. \& Rees, M. J. 1992, MNRAS, 255, 346

\reference Bartelmann, M., Scheider, P. 1993, A\&A, 268, 1

\reference Bartelmann, M., Scheider, P. \& Hasinger, G. 1994, A\&A, 290, 399

\reference Bogart, R. S. \& Wagoner, R. V. 1973, ApJ, 181, 609

\reference Boyle, B. J., Fong, R. \& Shanks, T. 1988, MNRAS, 231, 897

\reference Draine, B. T. \& Salpeter, E. E. 1979, ApJ, 231, 77

\reference Dwek, E., Rephaeli, Y. \& Mather, J. C. 1990, ApJ, 350, 104

\reference Fabian, A. C., Nulsen, P. E. J. \& Canizares, C. R. 1991, A\&ARv, 2, 191

\reference Fall, S. M. \& Pei, Y. C. 1993, ApJ, 402, 479

\reference Forman, W., Jones, C., Tucker, W. 1985, ApJ, 293, 102

\reference Giovanelli, R., Haynes, M. P., Salzer, J. J., Wegner, G., Da Costa,
L. N., 
Freudling, W. 1994, AJ, 107, 2036

\reference Goudfrooij, P., de Jong, T., Hansen, L. \& N\o rgaard-Nielsen, H. U.
1994, MNRAS, 271, 833

\reference Grevesse, N., \& Anders, E. 1991, Solar Interior \& Atmosphere, Tucson, AZ,
University of Arizona Press, 1227

\reference Heisler, J. \& Ostriker, J. P. 1988, ApJ, 332, 543

\reference Hu, E. M., Cowie, L. L. \& Wang, Z. 1985, ApJS, 59, 447

\reference Hu, E. M. 1992, ApJ, 391, 608

\reference Jansen, R. A., Knappen, J. H., Beckman, J. E., Peletier, R. F. \&
Hes, R. 1994, MNRAS, 270, 373

\reference Lees, J. F., Knapp, G. R., Rupen, M. P. \& Phillips, T. G. 1991, ApJ,
379, 177

\reference Lilly, S. J. \& Cowie, L. L. 1987, In Infrared Astronomy with
Arrays, eds. C. G. Wynn-Williams and E. E. Becklin, 473

\reference Margolis, S. H. \& Schramm, D. N. 1977, ApJ, 214, 339

\reference Masci, F. J. 1997, PhD thesis, University of Melbourne, astro-ph/9801181

\reference Masci, F. J., \& Webster, R. L. 1998, MNRAS, submitted

\reference McDowell, J. C. 1986, MNRAS, 223, 763

\reference Nath, B. B. \& Trentham, N. 1997, MNRAS, 291, 505

\reference Olive, K. A., Schramm, D. N., Steigman, G. \& Walker, T. P. 1990,
Phys.Lett.B, 236, 454

\reference Ostriker, J. P. \& Heisler, J. 1984, ApJ, 278, 1

\reference Pei, Y. C. 1992, ApJ, 395, 130

\reference Pei, Y. C. \& Fall, S. M. 1995, ApJ, 454, 69

\reference Peletier, R. F., Valentijn, E. A., Moorwood, A. F. M., 
Freudling, W., Knapen, J. H.
\& Beckman, J. E. 1995, A\&A, 300, L1

\reference Rodrigues-Williams, L. L. \& Hogan, C. J. 1994, AJ, 107, 451

\reference Rodrigues-Williams, L. L. \& Hawkins, C. J. 1995,
Proc. of the 5th Annual Astrophysics Conf.,
Maryland

\reference Romani, R. W. \& Maoz, D. 1992, ApJ, 386, 36

\reference Rudnicki, K. 1986, in Proc. Internat. of Physics ``Enrico Fermi'',
Summer School, 86, 480

\reference Sarazin, C. L. 1986, Rev. Mod. Phys., 58, 1

\reference Seitz, S. \& Scheider, P. 1995, A\&A, 302, 9

\reference Songaila, A. \& Cowie, L. L. 1996, AJ, 112, 335

\reference Takeda, H., Nulsen, P. E. J. \& Fabian, A. C. 1984, MNRAS, 208, 261

\reference Tinsley, B. M. 1976, ApJ, 208, 797

\reference Tyson, J. A. 1988, AJ, 96, 1

\reference V\'{e}ron-Cetty, M. P., \& V\'{e}ron, P. 1989, A Catalogue of Quasars
and Active Nuclei, 4th Edition, Munich: ESO

\reference Wang, B. 1991, ApJ, 374, 456

\reference Womble, D. S., Sargent, W. L. W., \& Lyons, R. S. 1996, in Cold Gas
at High Redshift, eds. M. Bremer, H. Rottgering, P. van der Werf,
C. Carilli, (Kluwer)

\reference Webster, R. L., Francis, P. J., Peterson, B. A., Drinkwater, M. J.,
\& Masci, F. J. 1995, Nature, 375, 469

\reference White, S. D. M., \& Frenk, C. S. 1991, ApJ, 379, 52

\reference Wise, M. W., O'connell, R. W., Bregman, J. N. \& Roberts, M. S. 1993,
ApJ, 405, 94

\reference Wright, E. L. 1990, ApJ, 353, 411

\reference Zaritsky, D. 1994, AJ, 108, 1619

\end{document}